\documentclass[]{scrartcl}

\usepackage[T1]{fontenc}
\usepackage[british]{babel}
\usepackage[utf8]{inputenc}
\usepackage[light, frenchstyle, narrowiints, partialup]{kpfonts}
\usepackage[scaled=0.92]{helvet}
\usepackage{courier}
\usepackage{amssymb,amsmath}
\usepackage[defaultlines=2,all]{nowidow}

\usepackage{graphicx}
\usepackage{authblk}

\title{Online survey for collective clustering of computer generated architectural floor~plans}

\author[1]{David~Sousa-Rodrigues\thanks{Corresponding author: {\tt david.rodrigues@open.ac.uk} (David Sousa-Rodrigues)\\ Extended abstract accepted for ICTPI’15 conference, June 17–19, Milton Keynes, United Kingdom – {\tt http://www.ictpi15.info/}}}
\author[2]{Mafalda~Teixeira~de~Sampayo}
\author[3]{Eugénio~Rodrigues}
\author[4]{Adélio~Rodrigues~Gaspar}
\author[5]{Álvaro~Gomes}
\author[5]{Carlos~Henggeler~Antunes}

\affil[1]{Centre of Complexity and Design, Faculty of Maths, Computing and Technology,\authorcr The Open University, Milton Keynes, United\,Kingdom}
\affil[2]{CIES, Department of Architecture, \authorcr Lisbon University Institute, Lisbon, Portugal}
\affil[3]{ADAI, LAETA, INESC Coimbra, Department of Mechanical Engineering,\authorcr University of Coimbra, Coimbra, Portugal}
\affil[4]{ADAI, LAETA, Department of Mechanical Engineering;\authorcr University of Coimbra, Coimbra, Portugal}
\affil[5]{INESC Coimbra, Department of Electrical and Computer Engineering,\authorcr University of Coimbra, Coimbra, Portugal}

\usepackage{hyperref} 
\hypersetup
{
    pdfauthor={David Sousa-Rodrigues, Mafalda Teixeira de Sampayo, Eugénio Rodrigues, Adélio Rodrigues Gaspar, Álvaro Gomes, and Carlos Henggeler Antunes},
    pdfsubject={The aim of this study is to understand what are the collective actions of architecture practitioners when grouping floor plan designs},
    pdftitle={Online survey for collective clustering of computer generated architectural floor plans},
    pdfkeywords= {Online Survey} {Generative Design} {Clustering} {Collective Intelligence} {Floor Plan Design} {Architecture} {Education}
}

\date{}                                           

\begin{document}
\maketitle
\noindent\hangindent=1.5em\textbf{Extended Abstract}

~

\noindent\hangindent=5.2em\textbf{Keywords:} Online Survey, Generative Design, Clustering, Collective Intelligence, Floor Plan Design, Architecture, Education.

~ 

The aim of this study is to understand what are the collective actions of architecture practitioners when grouping floor plan designs. The understanding of how professionals and students solve this complex problem may help to develop specific programmes for the teaching of architecture. In addition, the findings of this study can help in the development of query mechanisms for database retrieval of floor plans and the implementation of clustering mechanisms to aggregate floor plans resulting from generative design methods. The study aims to capture how practitioners define similarity between floor plans from a pool of available designs. A hybrid evolutionary strategy is used, which takes into account the building’s functional program to generate alternative floor plan designs [1–3].
The first step of this methodology consisted in an online survey to gather information on how the respondents would perform a clustering task. Online surveys have been used in several applications and are a method of data collection that conveys several advantages. When properly developed and implemented, a survey portrays the characteristics of large groups of respondents on a specific topic and allows assessing its representation. Several types of surveys are available; e.g. questionnaire and interview formats, phone survey, and online surveys, which can be coupled with inference engines that act and direct the survey according to respondents’ answers [4,5]. In the present study, the survey was posed as an online exercise in which respondents had to perform a pre-defined task, which makes it similar to running an experiment in an online environment. The experiment aimed to understand the perception and criteria of the target population to perform the clustering task by comparing the results with the respondents’ answers to a questionnaire presented at the end of the exercise.

\begin{figure}[htbp]
\centering
\includegraphics{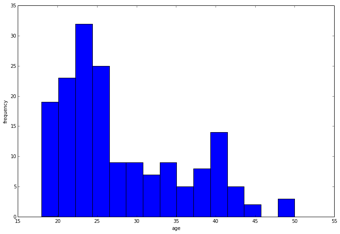}
\caption{Age distribution of respondents}
\label{fig1}
\end{figure}

The target group of this survey is individuals whose daily activities are related to architecture, i.e. architects, architecture students, civil engineers, and urban planners. The pool of participants inhabits mainly in Portugal and the ages range between 18 and 50 years old. Figure 1 depicts the age distribution of the respondents.

 \begin{figure}[htbp]
\centering
\includegraphics{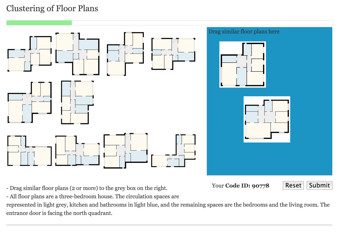}
\caption{Task Panel of the survey. Users must drag-and-drop to the blue area the floor plans presented on the left according to their notion of similarity.}
\label{fig1}
\end{figure}

The task was performed online through a web application. From a population of 72 floor plans, twelve randomly selected designs are chosen and displayed on screen. The user is then asked to drag-and-drop to a specific screen area the plans that he considers similar (see figure 2).
The 72 floor plans were generated using the Evolutionary Program for the Space Allocation Program (EPSAP) [1–3]. This algorithm is capable of producing alternative floor plans according to the same user’s preferences and requirements set as the functional program. This defines the type of building to be generated and the design constraints. The solutions generated were a single-family house with three bedrooms, one hall, one kitchen, a living room, one corridor and two bathrooms. A bathroom and all the bedrooms are connected to the corridor and all remaining spaces are connected to the hall. The kitchen also presents an internal door connecting it to the living room. One of the bathrooms serves the public areas of the house while the other connects to the corridor of the private area of the house. All inner rooms have doors of 90cm width, the exception being the living room doors that are 140cm. With the exception of the circulation areas and one of the bathrooms, all areas have at least one window — the living room has two. The hall has a door to the exterior facing North. No other restrictions were imposed on the functional program of this project. All solutions presented to the participants were previously generated and there was no human intervention in their selection for this exercise.
The participants were asked to perform an iterated task — ten times — of selecting similar floor plans. At the end of those ten iterations, a final form is presented for the respondent to identify the criteria used in the selection of the designs. At this moment the participant could also review — but not change — his previous selections. After submission the exercise was finished.
The data obtained were analysed after the construction of two square matrices — one representing in each entry the number of co-visualisation of the floor plans, i.e. the number of times floor plan i and floor plan j were shown in the same iteration; and the second matrix representing the number of co-selections of the floor plans by the user, i.e. the number of times the pair was selected as similar. The first matrix is the co-occurrence matrix while the second is the co-selection matrix.
A normalised matrix is constructed by division of the two previous matrices. The normalised matrix gives the fraction of times each pair of floor plans was selected. This matrix can be understood as an adjacency matrix where the entries represent the weights of the connections between two floor plan designs. The results present some background uncertainty and it is necessary to define a minimum threshold for the entries of the matrix. The value of the threshold is varied to identify the structure of the selection process. The resulting floor plan’s network represents the structure of the selection made by the participants. This network — undirected and weighted — is partitioned with the edge betweenness community detection algorithm by Girvan and Newman [6]. This is a divisive hierarchical algorithm that aims to find communities by maximizing the value of modularity — networks with high modularity have dense intra–cluster connections but sparse connections between vertices of different clusters. The graph and the resulting partition is characterised according to diverse properties — degree distribution, clustering coefficient, assortativity, small-world, and scale invariance. 

\begin{figure}[htbp]
\centering
\includegraphics[width=8cm]{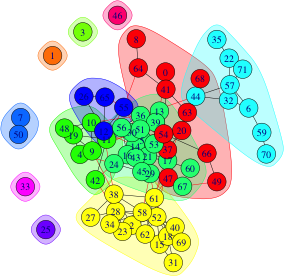}
\caption{Communities detected for the floor plans designs with threshold of 15\%}
\label{fig3}
\end{figure}

We show how topological properties emerge in the floor plan’s network, and characterise it by showing how the communities are identified by the collective answers of the respondents. In the case when no threshold is applied to the adjacency matrix the resulting network presents a single giant component with 15 complete cliques — subsets of vertices where the induced subgraph is complete, i.e. every two vertices are connected — and a network diameter of 2. When applying a 15
By performing a sweep of the threshold of the minimum percentage of selections, when two plans are shown in common, it is possible to identify the floor plans that are the roots of the different typologies. The sets of floor plans are not defined in a hierarchical manner but some pairs of floor plans will naturally be co-selected more often than others. Thus hierarchies of pairs of floor plans based are defined on their co-selection frequency. 
The understanding of how people perform certain tasks is crucial for the development of education strategies focused on improving learning at university level education. Several studies have been proposed that include the participation of the crowd and are bottom up learning processes, e.g. peer assessment [7] where students mark each other’s work. In this study the process of grouping floor plans is investigated to understand the criteria used by the students and other practitioners. The results are presented and discussed in light of teaching strategies for the architecture education at the university level. The results show how collective action on simple tasks can lead to the emergence of the solution for the complex task of defining hierarchies of similarity in floor plan’s designs and identifying the criteria used by a class of professionals.
The results obtained in this work are important for future development of ICT-mediated strategies for architecture education and professional practitioners. They will also impact other applications such as floor plan design database retrieval and aggregation of similar solutions that result from generative design methods. The criteria reported by the respondents varied and can be incorporated in machine learning algorithms to perform the clustering tasks presented to humans in ways that mimic experts’ actions. 

\vfill

\noindent\textbf{References} \\

\noindent\hangindent=1.5em
[1] Rodrigues E, Gaspar A, Gomes Á. An evolutionary strategy enhanced with a local search technique for the space allocation problem in architecture, Part 1: Methodology. Computer Aided-Design. 2013;45(5):887-- 897.

\noindent\hangindent=1.5em
[2] Rodrigues E, Gaspar A, Gomes Á. An evolutionary strategy enhanced with a local search technique for the space allocation problem in architecture, Part 2: Validation and Performance Tests. Computer Aided-Design. 2013;45(5):898--910.

\noindent\hangindent=1.5em
[3] Rodrigues E, Gaspar A, Gomes Á. An approach to the multi-level space allocation problem in architecture using a hybrid evolutionary technique. Automation in Construction. 2013 November;35:482--498.

\noindent\hangindent=1.5em
[4] Urbano P, Sousa-Rodrigues D. Rule Based Systems Applied To Online Surveys. In: IADIS WWW/Internet Conference. Freiburg; 2008.

\noindent\hangindent=1.5em
[5] Urbano P, Sousa-Rodrigues D. The Advantage Of Using Rules in Online Surveys. Revista de Ciências da Computação. 2008;III(3).

\noindent\hangindent=1.5em
[6] Girvan M, Newman MEJ. Community structure in social and biological networks. Proceedings of the National Academy of Sciences. 2002;99(12):7821--7826.

\noindent\hangindent=1.5em
[7] de Sampayo MT, Sousa-Rodrigues D, Jimenez-Romero C, Johnson JH. Peer Assessment in Architecture Education. In: International Conference on Technology and Innovation. Brno, Czech Republic; 2014.

\end{document}